# Photonic crystal dumbbell resonators in silicon and aluminum nitride integrated optical circuits

W. H. P. Pernice[1,2], Chi Xiong[1] and H. X. Tang[1*]

[1]Department of Electrical Engineering, Yale University, New Haven, CT 06511, USA

[2]Current address: Institute of Nanotechnology, Karlsruhe Institute of Technology (KIT), 76344 Karlsruhe, Germany

**Abstract**: Tight confinement of light in photonic cavities provides an efficient template for the realization of high optical intensity with strong field gradients. Here we present such a nanoscale resonator device based on a one-dimensional photonic crystal slot cavity. Our design allows for realizing highly localized optical modes with theoretically predicted Q factors in excess of $10^6$. The design is demonstrated experimentally both in a high-contrast refractive index system (silicon), as well as in medium refractive index contrast devices made from aluminum nitride. We achieve extinction ratio of 21dB in critically coupled resonators using an on-chip readout platform with loaded *Q* factors up to 33,000. Our approach holds promise for realizing ultra-small opto-mechanical resonators for high-frequency operation and sensing applications.

---

[*] Email: hong.tang@yale.edu

Over the last decade considerable research effort has been devoted to developing ultrahigh quality factor ($Q$ factor) photonic crystal (PhC) cavities that have dimensions comparable to the wavelength of light [1-4]. By shrinking the modal volume to near the wavelength size $V=(\lambda/2n)^3$, these cavities have enabled different applications to emerge in ultrasmall lasers [5,6], strong light-matter interactions [7-9] as well as optical switching [10] among others. Besides designs in two-dimensional photonic crystal cavities, there has been much interest in cavities realized in suspended nanobeams patterned with a one-dimensional lattice of holes [11–14]. 1D PhC cavities comprise exceptional cavity figures of merit ($Q/V$), relative ease of design and fabrication, and show much potential for the exploitation of optomechanical effects [15,16]. While such cavity designs have been investigated in stand-alone configurations in which the cavity is read out via free-space optical setups or fiber tapers, their true potential lies in the integratability with traditional nano-photonic circuitry. Even though PhC cavities provide exceptionally high optical $Q$, even smaller mode volumes can be realized in optical cavity geometries that contain slot waveguides [17,18]. Because the electromagnetic mode is highly confined in the thin air gap between adjacent cladding structures, the modal volume is defined by the dimensions of the air slot which is only limited by achievable fabrication tolerances. Such a design is attractive for optomechanical applications, because strong optical field gradients can be achieved which result in strong gradient optical forces.

Here we design and demonstrate a high $Q$ one-dimensional photonic crystal cavity combined with a central slot region, which we term a PhC dumbbell cavity. The name is chosen, because the cavity area defined predominantly between the two inner most holes connected by the rectangular slot section resembles a dumbbell weight. By employing tapered Bragg mirrors alongside the air slot we achieve low loss optical confinement of the cavity fields into the slot region. The cavity is integrated into an on-chip photonic circuit and coupled to a feeding waveguide for convenient optical access. In this design strong field enhancement occurs in the slot area, as a prerequisite for high efficiency optomechanical coupling.

We first investigate the cavity design numerically, using the finite-difference time-domain method. As illustrated in Fig.1(a) we employ a cavity framework, in which the slot region is enclosed between linearly tapered PhC Bragg mirrors. In this design the radii of the holes are continuously decreased from a larger radius in the unperturbed photonic crystal region towards a small radius in the cavity area. For the high refractive index contrast case we assume that the cavity is fabricated from a free-standing silicon beam with refractive index of 3.48 and a height of 220 nm. The relevant optimization parameters as shown in Fig.1(a) are the length of the taper, the hole radius at the cavity center $R_2$ and in the mirror region $R_1$, the waveguide width and the dimensions of the slot. The hole radius is defined as the lattice constant $a$ times a fixed filling factor $f$ as $R=f*a$. The filling factor is optimized to be 0.29. By scanning the multi-dimensional parameter space we arrive at an optimized design space where the cavity degrees of freedom are: $R_1=f*440$ nm, $R_2=f*300$ nm, $w=500$ nm and the number of tapered holes is 5. The radii of the two holes making up the dumbbell structure are chosen to be the same as $R_2$. With this parameter set we obtain maximum $Q$ factors in excess of $10^6$ as shown in Fig.1(b). The $Q$ factors depend here on the slot width and the length of the slot. When the slot length is varied, the cavity resonance increases linearly with decreasing slot length. Our best $Q$ factors are predicted for a slot length of 185 nm and a slot width of 100 nm, where the highest $Q$ exceeds $2\times10^6$. For this geometry we numerically extract the normalized modal volume by evaluating the integral [19] $V_{eff} = \dfrac{\int \varepsilon(\vec{r})|\vec{E}(\vec{r})|^2 d^3r}{\varepsilon(\vec{r}_{max})\max\left\{|\vec{E}(\vec{r})|^2\right\}}\left(\dfrac{2n(\vec{r}_{max})}{\lambda}\right)$, where $\vec{r}_{max}$ is the location of the maximum squared electric field and $\lambda$ is the resonance wavelength. Thus the volume can be reduced by increasing the mode maximum electric field and localizing the mode maximum in the low index region. Reducing $V_{eff}$ in cavities enables control of the degree of light-matter interaction for processes such as nonlinear optics and cavity quantum electrodynamics. Because in the dumbbell design the maximum field is located in the low refractive index region (air in our case), the mode volume is reduced compared to traditional dielectric cavities. For our silicon design we calculate effective mode volume of $0.06(\lambda/n_{air})^3$, in agreement with previous results [18-20].

To confirm the optical properties of designed structures, integrated photonic devices are fabricated from silicon-on-insulator (SOI) substrates with a 220 nm thick silicon layer on top of a 3μm thick buried oxide layer. Electron-beam lithography using ZEP520A positive tone resist and subsequent reactive ion etching in chlorine chemistry are used to define the cavity and supporting on-chip photonic circuitry. After dry etching a second *e*-beam lithography step with ZEP520A is performed to define opening windows for the subsequent release of the cavity beam. A timed wet etching step with buffered oxide etchant (BOE) is then used to selectively remove the oxide layer in the release windows under the cavity beam. A fabricated device is shown in the optical microscope image in Fig.2(a). Grating couplers are used to launch light into the feeding waveguide, which is routed to the cavity as shown in the dark field image zoom into the coupling region for coupling to the feeding waveguide. The wet release leads to free-standing feeding waveguides and cavity beams as shown in the SEM picture inset Fig.2(a). Due to the anisotropic nature of the wet etching process, sloped sidewalls towards the edges of the nanobeams result. The cavity and the feeding waveguide are released to a depth of about 1μm, leading to a sufficient undercut of the cavity beam so that the influence of the substrate is minimized.

The optical properties of the fabricated device are then investigated by measuring the transmission through the feeding waveguide. In order to optimize the coupling condition in the cavity we fabricate a variety of structures with varying coupling gap between the cavity beam and the feeding waveguide. Optimal coupling conditions are found for a coupling gap of 180nm, leading to near critical coupling with high extinction ratio. We use a tunable diode laser (New Focus 6428) to scan the input wavelength across the cavity resonance as shown in Fig.3(a). The transmission profile shows the envelope of the grating coupler, with small fringes due to back-reflection from the output grating coupler. A clear resonance dip indicates the cavity wavelength at 1552.8 nm. By fitting the resonance with a Lorentzian as shown in Fig.3(b), the quality factor of the cavity can be extracted. For best devices as shown in Fig.3(b) we find a loaded cavity Q of 20,000. When the input waveguide is added the intrinsic Q factor simulated in Fig.1(b) can be significantly compromised. For the experimentally optimized coupling gap of 180nm we expect a cavity Q of 340,000 from additional FDTD simulations. The experimentally observed quality factor is reduced compared to the simulation value due to the presence of the substrate in the evanescent cavity field and the surface roughness due to fabrication imperfections.

As expected from the FDTD simulations described above we expect the cavity resonance to depend on the central slot length. Therefore we fabricate a number of devices with slot lengths ranging from 160 nm to 220 nm. For each slot length we obtain the cavity resonance at the point of critical coupling as shown in Fig.3(b). Therefore for each slot length we also fabricate a number of devices with varying coupling strength, to make sure that the cavity resonance is measured under equal experimental conditions. For the shortest slot length of 160 nm we find a cavity resonance at 1548 nm. By increasing the slot length to 220 nm the cavity resonance shifts by 4.8 nm to 1552.8 nm. The results are compared against FDTD simulations shown by the red markers in Fig.3(b). Good agreement between the two is found for the simulated geometry.

While silicon provides high refractive index contrast and allows for sub-wavelength optical components, the small bandgap of only 1.1eV restricts the use of silicon to wavelengths above 1100 nm. Furthermore, silicon is plagued by strong free-carrier absorption, which implies that the strong field enhancement in nanoscale cavities leads to thermal instability if the input optical power is high. As a result silicon nanobeam cavities have to be operated at relatively low optical driving power, partially cancelling the expected benefit from high cavity optical $Q$. Therefore alternative material systems with larger bandgaps are attracting increasing attention, because they allow for the realization of photonic components that can be operated in the visible wavelength range and also at high optical input powers. One of the largest bandgaps is offered by aluminum nitride, which has recently been demonstrated to be a versatile material for integrated photonic circuits [21-23]. Aluminum nitride supports waveguiding down to 220 nm. However, due to a smaller refractive index of 2.10, light is less strongly confined compared to silicon devices and therefore a modified dumbbell design is employed to obtain high optical quality factors numerically. Here instead of tapering the hole radius we vary the lattice constant of the photonic crystal from a larger period $l_1$=580nm to a smaller period $l_2$=440nm in the cavity region. Differing from the silicon design, the tapering process in AlN devices is done parabolically to provide a near adiabatic transition between the PhC lattice and the cavity. In this case we arrive at comparable optical quality factors above 1 million for the isolated cavity design. In the AlN design we calculate effective mode volumes of $0.11(\lambda/n_{air})^3$, which are larger compared to the silicon design because of a smaller refractive index contrast.

Simulated optimal devices are fabricated using an approach modified from the silicon routine described above. As in previous work we use a self-protecting masking approach performed with a two-step etching procedure. For AlN we use positive tone HSQ resist and dry etching in Ar/$Cl_2$/$BCl_3$ chemistry. In the first step the photonic circuitry is etched down to a residual depth of 70nm. Then a second e-beam lithography step in ZEP520A resist is performed to define the opening windows. A second dry etch is used to remove the residual AlN in the opening areas, while the waveguide and cavity beams are still protected by the HSQ resist remaining on the photonic circuits. A final wet etching in BOE is used to release the beams. In this case the residual AlN layer provides a natural mask against the BOE and also defines precise clamping points of the nanobeams. Fabricated circuits are shown in the optical image in Fig.2(b), again with grating coupling input ports, waveguides and cavity beams. The cavity area is underneath the alignment marker structure in the zoom image in Fig.2(b). Here we provide weak input coupling by using point coupling with an arc design. On the center of the beam the dumbbell slot region is clearly visible in the close-up SEM image in Fig.2(b).

As in the case of silicon dumbbell devices we vary the cavity geometry and coupling conditions across fabricated chips of many devices. For each device the transmission characteristics are recorded and the properties of the cavity resonance explored. As shown in Fig.4(a) we find best optical $Q$ for a device with a slot length of 700 nm, a slot width of 100nm and lattice constants of the outer photonic crystal area of 560 nm, while the central lattice constant is 430nm. The filling factor is again kept at 0.29. For a beam width of 950 nm we measure optical Q of 33,000 at a wavelength of 1534.1 nm for a weakly coupled device. By varying the distance between the point contact and the cavity beam we can tune the coupling conditions from the overcoupled regime into the undercoupled regime, as shown in Fig.4(b). For near critical coupling we obtain extinction ratio above 20 dB, comparable to values measured on optical ring resonators. As in the silicon devices the cavity resonance can be tuned to desired wavelengths by varying the length of the slot region. However, differing from the silicon architecture free-carrier based absorption effects are not found during the measurements, illustrating that AlN allows for higher optical input power. Furthermore, the lower refractive index of 2.10 at 1550 nm provides an optical mode that is less confined within the waveguide material, which may be a reason for the slightly higher quality factors obtained in the measurements.

In conclusion we have experimentally demonstrated a dumbbell shaped slot cavity system with ultra-small mode volume and high quality factor. Both in silicon-on-insulator substrates with high refractive contrast and low-refractive index aluminum nitride, we achieve optical resonances in the telecoms C-band. By controlling the coupling distance into the cavity region we show high extinction ratio in excess of 20 dB. Furthermore, the cavity resonance can be broadly tuned by varying the length of the central slot. Our design provides highly concentrated fields in a small special volume, leading to large gradients of the electrical fields. In combination with high optical quality factors measured above 30,000 the dumbbell geometry will be of interest for the study of optomechanical interactions as well as for experiments for optical particle trapping.

This work was supported by a seedling program from DARPA/MTO and the DARPA/MTO ORCHID program through a grant from AFOSR. W.H.P. Pernice would like to thank the Alexander-von-Humboldt foundation for providing a postdoctoral fellowship. Facilities use was supported by YINQE and NSF MRSEC DMR 1119826. H.X.T acknowledges support from a Packard Fellowship in Science and Engineering and a CAREER award from the National Science Foundation.

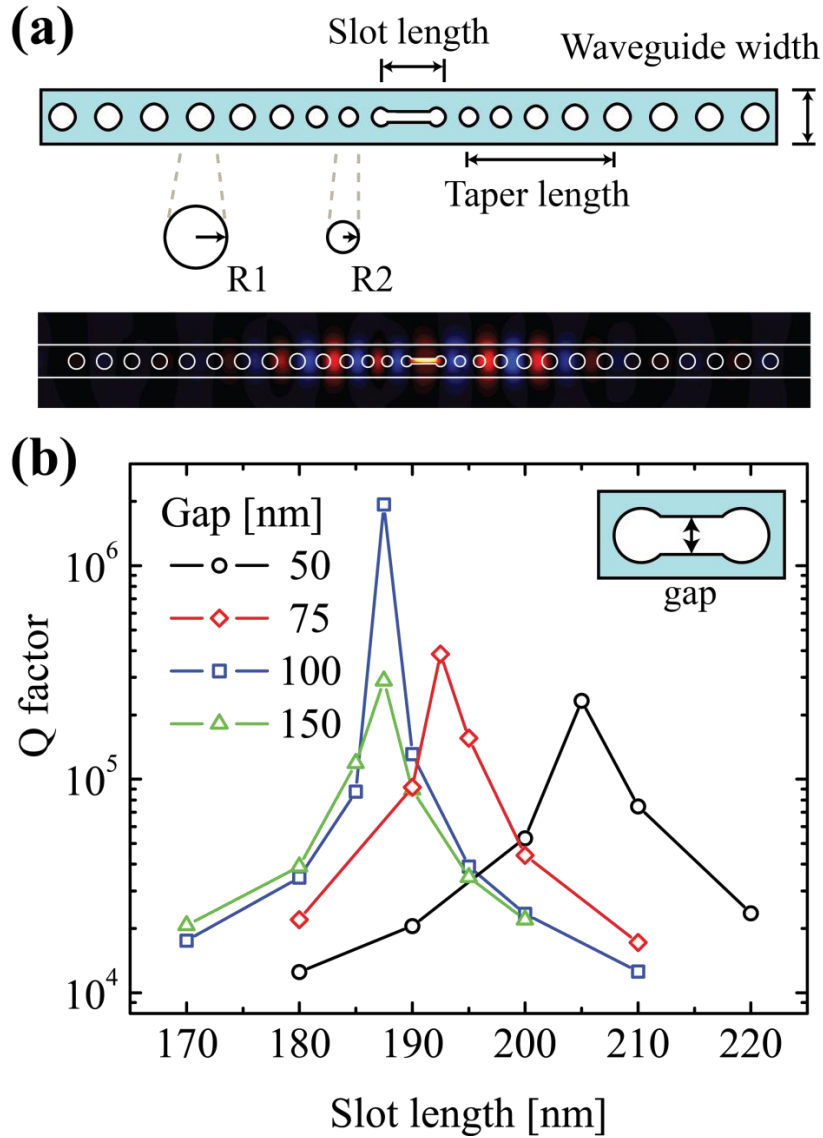

**Figure 1.** (a) The design of the photonic crystal dumbbell slot cavity. The hole radius is tapered from $R_1$ to $R_2$, while the inner holes are connected by a narrow air gap. The simulated optical cavity mode shows highly concentrated electrical fields in the slot region. (b) The simulated intrinsic optical $Q$ factor as a function of the slot length and the slot width for a silicon nanobeam. Best cavity $Q$ of 2 million is predicted for a slot width of 100nm.

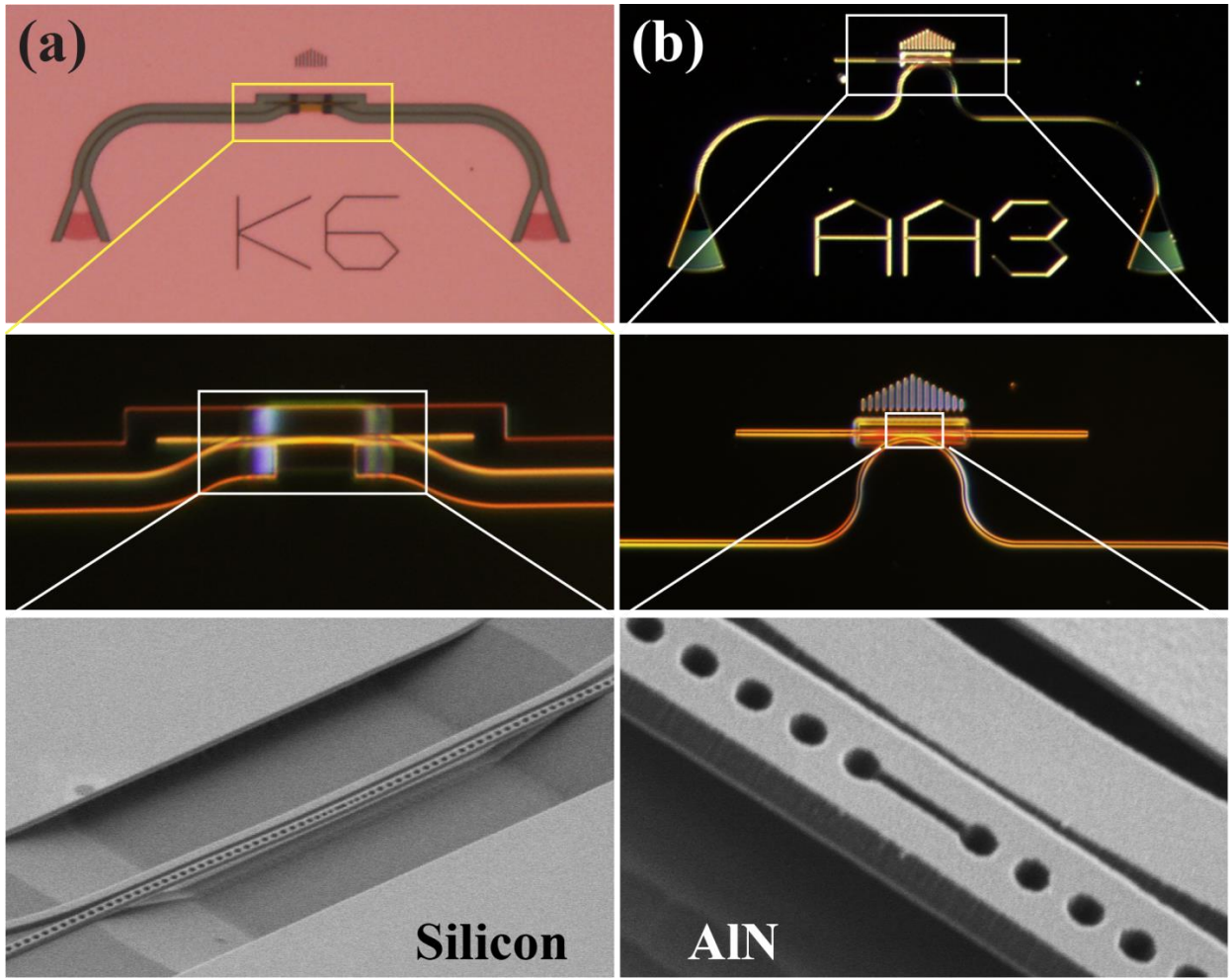

**Figure 2.** (a) A fabricated photonic circuit for the measurement of side-coupled dumbbell cavities in silicon. Inset: A darkfield optical image of the coupling region between the feeding waveguide and the nanobeam, with a SEM picture of the released section of the photonic circuit. (b) A corresponding circuit for the measurement of AlN dumbbell devices. Inset: point coupling geometry and a SEM image of the dumbbell region.

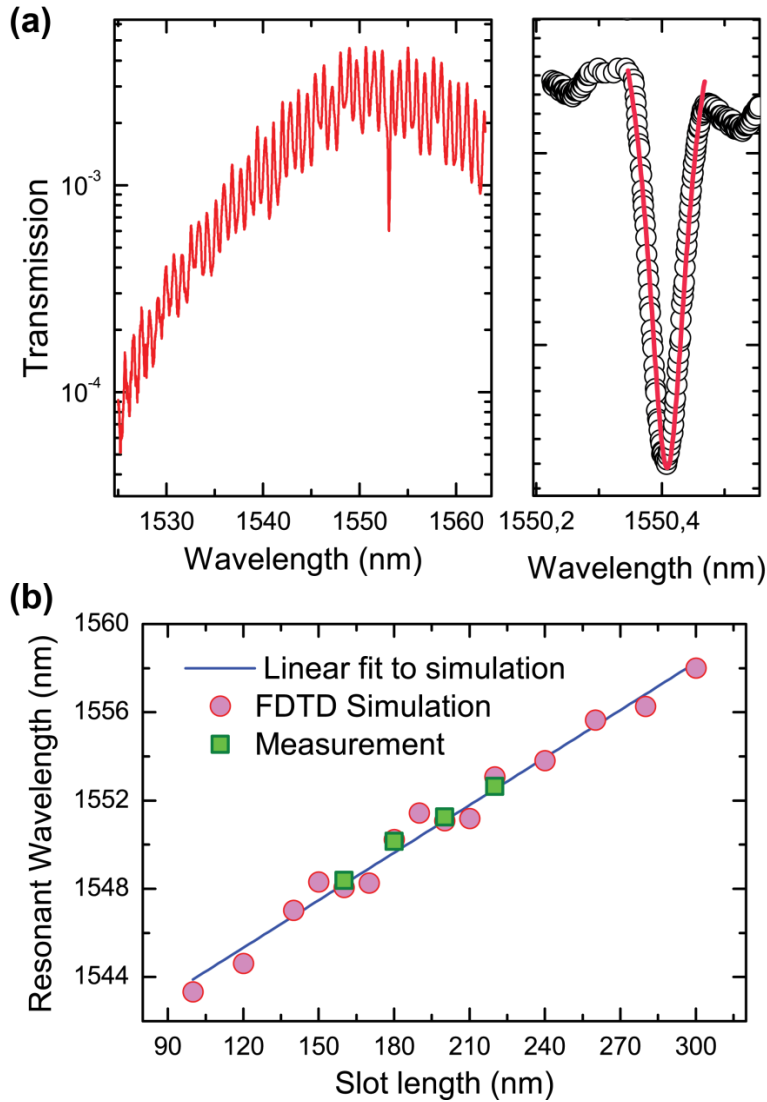

**Figure 3**. (a) The measured transmission spectrum of a fabricated silicon device, showing the cavity resonance at 1552.8 nm. The Lorentzian fit to one of the sharpest resonances at 1550.4 nm reveals an optical Q of 20,000. (b) The measured optical response in dependence of the slot length. Shown are results from FDTD simulations (red markers) as well as measured data (green markers). The blue line is a linear fit to the FDTD data.

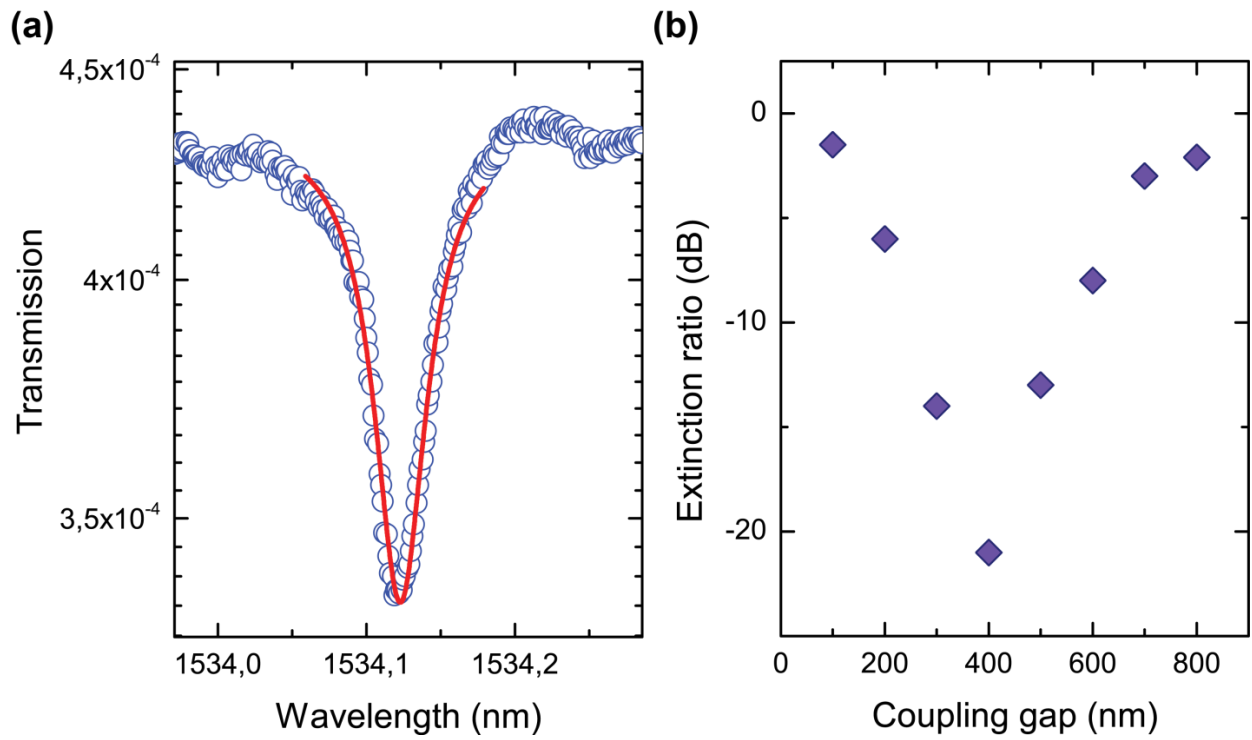

**Figure 4**. (a) Measured results for the AlN dumbbell cavity. Best optical Q of 33,000 is found for a device with a slot length of 700 nm and a slot gap of 100nm. (b) For critically coupled devices we find extinction ratio better than 20 dB.